\documentclass[11pt]{article}
\usepackage[margin=1in]{geometry}
\usepackage{amsmath}
\usepackage{amssymb}
\usepackage{graphicx}
\usepackage{lineno}
\usepackage{url}

\usepackage{xcolor}
\usepackage[normalem]{ulem}

\begin{document}

\title{A compact low-frequency optomechanical triaxial inertial sensor}

\author{%
\begin{minipage}{\textwidth}
\centering\footnotesize
Daniel George$^{1,\dagger,\ddagger}$,
Jose D. Hernandez Rivero$^{1,\dagger,*}$,
Moritz Mehmet$^{1,\S}$,
Ramses Miranda Espino$^{1}$,
Xiangyu Guo$^{1,\P}$,
Andrea Nelson$^{1,\|}$,
Jose Sanjuan$^{1}$,
and Felipe Guzman$^{1}$\\[6pt]
$^{1}$Wyant College of Optical Sciences, University of Arizona, Tucson, AZ 85721, USA\\[6pt]
$^{\dagger}$These authors contributed equally to this work.\\
$^{\ddagger}$dongeorge@arizona.edu \quad
$^{*}$jhernandezrivero@arizona.edu\\
$^{\S}$Present address: Leibniz University Hannover, 30167 Hannover, Germany, and Max Planck Institute for Gravitational Physics (Albert Einstein Institute), 30167 Hannover, Germany\\
$^{\P}$Present address: Arizona State University, School of Manufacturing Systems and Networks, Mesa, AZ, USA\\
$^{\|}$Present address: Lawrence Livermore National Laboratory, Livermore, CA, USA
\end{minipage}
}

\maketitle

\begin{abstract}
Triaxial optomechanical accelerometers offer compact, ground-testable alternatives to electrostatic sensors for satellite geodesy, seismometry, and various other applications. We demonstrate a low-CSWaP triaxial sensor using monolithic fused-silica resonators with dual heterodyne interferometric readout. The X axis reaches a 60~pico-g/$\sqrt{\mathrm{Hz}}$ readout noise floor, and the in-plane axes resolve ambient seismic ground motion in agreement with a co-located commercial seismometer from 4~mHz to 8~Hz. The Z axis exhibits higher noise under $1g$ loading due to gravity-induced geometric stiffening (44.45~Hz versus 11.07~Hz in a 0g-equivalent configuration), with measurements indicating in-orbit performance on par with the in-plane sensors. These results support the feasibility of triaxial optomechanical accelerometry for space missions and ground-based applications.
\end{abstract}


\section{Introduction}

High sensitivity inertial measurements at frequencies in the range of Hz down to mHz are crucial across a broad range of applications in satellite geodesy, drag-free control, inertial navigation, space based interferometry, and seismometry \cite{Sanjuan2023,Jennrich_2009,Conklin2015,rs14133092}. For geodetic applications, accurate measurement of non-gravitational accelerations is essential for refining weak gravitational signatures by isolating external disturbances. Missions such as the Gravity Recovery and Climate Experiment (GRACE) and its successor, the Gravity Recovery and Climate Experiment Follow-On (GRACE-FO), employ three-axis electrostatic accelerometers as witness sensors to quantify atmospheric drag, solar radiation pressure, and other perturbations. Subtraction of these effects in post-processing allows for a better recovery of Earth's gravity field and its temporal variations \cite{Tapley2004,Kornfeld2019,Flury2008,Bandikova2019}.

The sensing approach explored in this work is not limited to geodesy; it extends readily to a wider class of low-frequency inertial measurement problems, on ground and in orbit. We present a measurement of low-frequency seismic ground motion in the microseism band~\cite{LonguetHiggins1950, Ardhuin2011}, recorded in Tucson, Arizona, detected through direct deflections of the test masses in our optomechanical inertial sensor and cross-validated against a reference Trillium Horizon 120 seismometer installed on the same platform as our sensor.

The optomechanical accelerometers utilize optical interferometry to read out displacement, converting the motion of a mechanically suspended test mass into an acceleration signal. By combining low-loss mechanical resonators with laser-based displacement readout, such sensors provide SI-traceable measurements, reduced susceptibility to electromagnetic interference, and a compact footprint suitable for laboratory and field environments, through a reduced cost, size, weight, and power (CSWaP) system design. Low-frequency optomechanical resonators with $mQ$-products at levels of 250~kg can have projected acceleration noise floors on the order of $5\times10^{-11}\,\mathrm{m/s^2/\sqrt{Hz}}$ at 1~Hz with resonant frequencies below 10~Hz~\cite{Hines2022}, positioning them as promising candidates for future gravity field recovery and drag-free satellite missions.

Earlier single-axis optomechanical accelerometers developed in our laboratory have demonstrated low noise and strong low-frequency sensitivity, establishing the viability of optical readout for precision inertial sensing \cite{Hines2020}. This level of performance makes such devices promising candidates for seismometry and related ground-motion sensing applications---for example, the seismic isolation and monitoring systems of gravitational-wave observatories such as the Laser Interferometer Gravitational-Wave Observatory (LIGO), where precise detection of seismic accelerations is required for active suspension control systems. Furthermore, extending optomechanical inertial sensing to three orthogonal axes within a single instrument enables full vector acceleration measurements, which are essential for spacecraft dynamics, platform stabilization, and hybrid inertial navigation systems when combined with gyroscopes and attitude sensors \cite{Sanjuan2023}.

Complementary efforts by other groups have pursued compact mechanical sensing across a range of platforms. A MEMS gravimeter has resolved Earth tides over a continuous 19-day measurement, demonstrating that chip-scale devices can achieve the long-term stability needed for geophysical observation \cite{Prasad2022}. More recently, integrated photonic approaches have produced monolithically fabricated silicon tri-axial accelerometer chips capable of resolving 20--30 nano-g level accelerations over bandwidths of 1--100~Hz \cite{Hong2025}. These results illustrate the complementary nature of on-chip and bulk optomechanical sensing strategies.

In this work, we present a complete triaxial optomechanical accelerometer system designed for low-frequency inertial sensing. The system exhibits two in-plane axes with high laboratory sensitivity of approximately 60~pico-g$/\sqrt{\mathrm{Hz}}$ at 1~Hz and a third axis whose performance is currently limited by static gravitational loading and gravity-induced geometric nonlinearity. The behavior of this less sensitive Z-axis resonator is understood and motivates future microgravity testing to isolate intrinsic acceleration performance. The demonstrated performance, combined with the low CSWaP, establishes the feasibility of triaxial optomechanical acceleration measurement and a clear path toward flight-representative instrumentation for satellite geodesy, inertial navigation, and a myriad of precision optical metrology applications.

\section{Methodology}
\label{sec:methodology}

\subsection{Monolithic fused-silica resonators}
\label{sec:resonators}

At the heart of each inertial sensor in the triaxial system is an optomechanical resonator, which consists of a test mass suspended by thin flexures, etched as a single monolithic structure from a fused-silica plate. In the linear oscillation regime, the test mass displacement $\delta x(f)$ responds to an inertial acceleration $\delta a(f)$ according to the transfer function of a damped harmonic oscillator,
\begin{equation}
\frac{\delta x(f)}{\delta a(f)} = \frac{-1}{(2\pi f)^2 - (2\pi f_0)^2 + i (2\pi f)(2\pi f_0)/Q},
\label{eq:transfer_function}
\end{equation}
where $f_0$ is the resonant frequency and $Q$ is the mechanical quality factor. For $f \ll f_0$, Eq.~\eqref{eq:transfer_function} reduces to $\delta x(f) \approx \delta a(f)/(2\pi f_0)^2$; we choose low frequency oscillators to maximize this response and improve sensitivity against a fixed displacement readout noise floor.

The fused-silica wafers were fabricated by Femtoprint (Muzzano, Switzerland) from our design, using selective laser-induced etching (SLE), which enables high-aspect-ratio monolithic flexure structures with minimal subsurface damage relative to conventional wet or dry etching.

\begin{figure}[htbp]
  \centering
  \includegraphics[width=\linewidth]{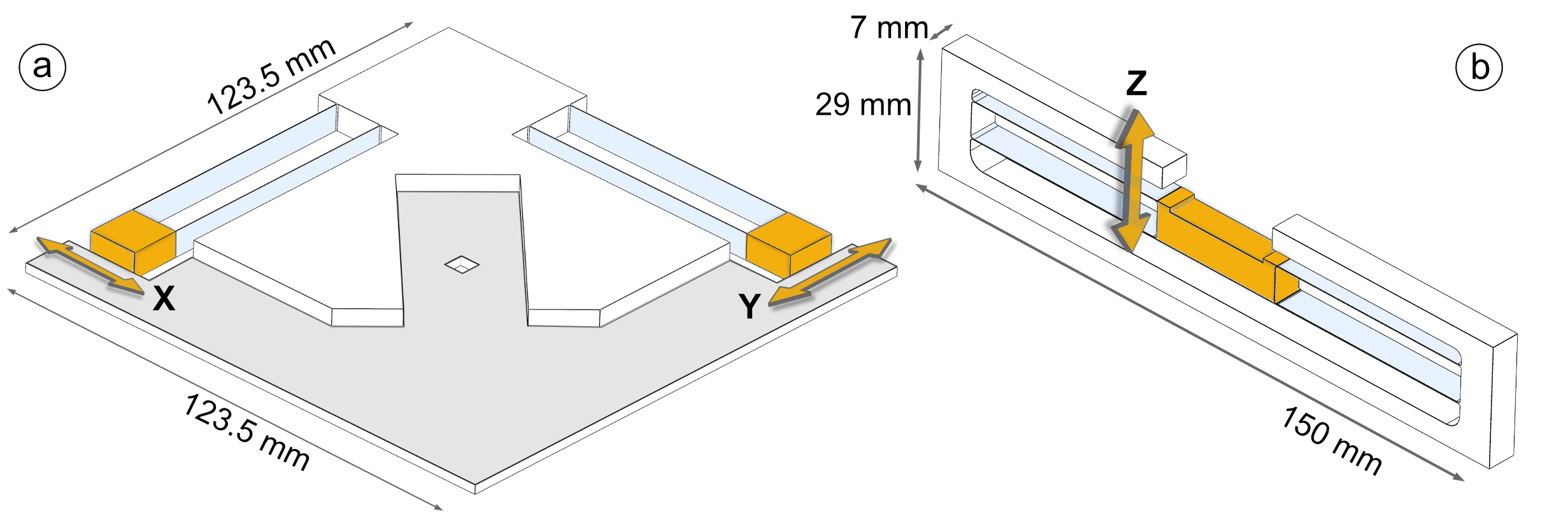}
  \caption{Layout of the mechanical resonators for sensing along the X and Y axes (a) and the Z axis (b). Both sub-systems are manufactured from 7\,mm thick fused-silica wafers. The test mass (TM) of each resonator is highlighted in yellow. TMs are supported by a set of 100\,$\mathrm{\mu}$m thin flexures (light blue). Arrows indicate the direction of TM motion. The gray part of (a) is used to integrate the interferometric readout for all axes, and the cut-out in the center of the plate allows us to steer a beam downwards to the Z resonator.}
  \label{fig:IMG1}
\end{figure}

\begin{figure}[htbp]
  \centering
  \includegraphics[width=\linewidth]{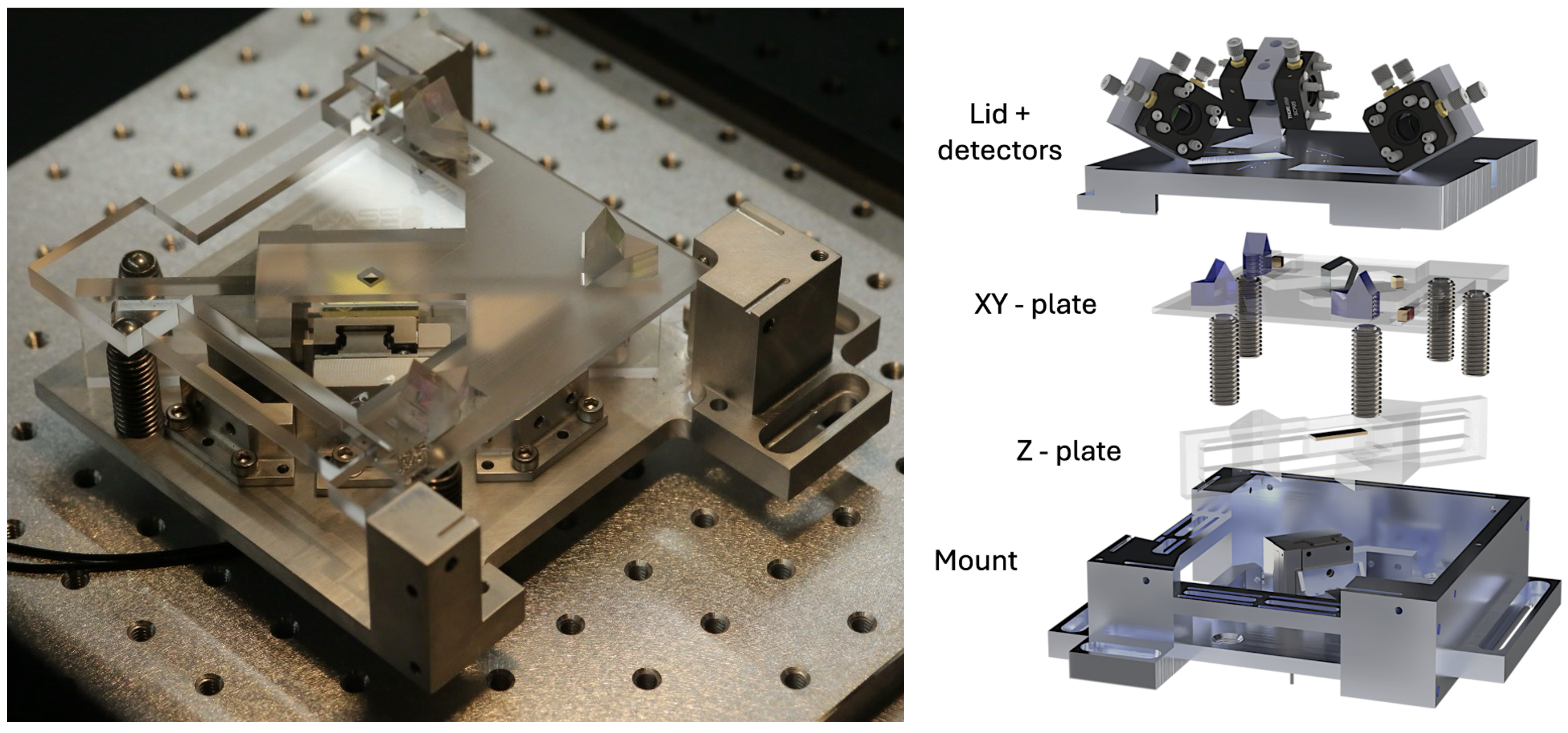}
  \caption{Image of the triaxial sensor, with an exploded render of the X, Y, and Z plates, mount, and conceptualized lid with self-contained readout optics and detectors.}
  \label{fig:CAD_XYZ}
\end{figure}

Operation in the three axes is sensed with resonators etched into two fused-silica plates, as shown in Fig.~\ref{fig:IMG1}(a,\,b). These sensors are pictured in the exploded view of Fig.~\ref{fig:CAD_XYZ}, where they are assembled and secured within an optomechanical mount. The Z-axis resonator plate sits on the base, supported by Invar spacers. The test mass is located in the center of the Z plate frame, suspended by four flexures in a bridge-like geometry. The XY plate incorporates two identical resonators close to two edges of the frame as shown in Fig.~\ref{fig:IMG1}(a). The X- and Y-axis oscillators follow a parallelogram design, inspired by our lowest-frequency sensor~\cite{Hines2022}. The mount contains set-screws in the base for the XY plate to sit, while also allowing for a coarse alignment of the interferometer with the input optical axes. Moreover, the interferometric optics for all three axes are housed on the wafer itself, allowing for a considerably compact optomechanical sensor. Together, this assembly forms the science sensor of the triaxial accelerometer, occupying a compact footprint of 156$\times$156$\times$65\,mm$^3$.

Resonant frequencies and quality factors for all three axes were determined from ring-down measurements in vacuum: each resonator was excited with an applied impulse and the free decay envelope
of oscillation amplitude was fit to an exponentially decaying sinusoid, from which $f_0$ and $Q$ were extracted. For the Z resonator, each measurement was repeated five times, and we report the sample mean as the central value and the sample standard deviation as the associated uncertainty. Uncertainties quoted for the in-plane resonators are estimated from the reproducibility of repeated ring-down characterizations of a nominally identical XY resonator at $p \lesssim 7\times10^{-4}$\,mbar.

The XY resonant frequencies and quality factors are summarized in Table~\ref{tab:xy_resonators}, where the resonant frequencies are below 7~Hz. Higher order modes are at least a factor of 20 higher in frequency according to finite-element simulations, ensuring minimal modal cross-talk. In air, gas damping dominates the mechanical loss, lowering the operational quality factor to a few thousand, and subsequently raising the thermal noise floor. The vacuum quality factors correspond to an $mQ$-product of roughly 400\,kg, where internal material losses (bulk, surface, and thermoelastic) dominate. The power spectral density of the thermal acceleration $S_{a_{\mathrm{th}}}$ can be approximated as~\cite{Hines2020,Hines2022,PhysRevD.42.2437}

\begin{equation}
S_{a_{\mathrm{th}}}(f) = \frac{4k_B T 2\pi {f_0}^2}{m Q f},
\label{eq:thermal_noise}
\end{equation}
where $k_B$ is Boltzmann's constant, $T$ is temperature in Kelvin, and $m$ is the mass of the oscillator. Detailed derivations for the contributing mechanical loss mechanisms can be found in~\cite{Hines2022}.

\begin{table}[htbp]
\caption{Resonant Frequencies and Quality Factors of the In-Plane Resonators}
\label{tab:xy_resonators}
\centering
\begin{tabular}{lcc}
\hline
Axis & $f_0$ (Hz) & $Q~(\times10^5)$ \\
\hline
X & $6.8845 \pm 0.0002$ & $2.07 \pm 0.30$ \\
Y & $6.8943 \pm 0.0002$ & $1.69 \pm 0.30$ \\
\hline
\end{tabular}
\end{table}

The fabricated Z-axis resonator exhibits more complex behavior, and is presented in Table~\ref{tab:z_resonator_summary}. Its resonant frequency increases from $f_{0,z}^{\mathrm{orbit}} = 11.07$\,Hz under near-zero static load to $f_{0,z}^{\mathrm{ground}} = 44.45$\,Hz under $1g$ load, due to gravity-induced flexure stiffening. To study the in-orbit performance, the Z-axis resonator was placed in a dedicated fixture oriented such that gravity is now perpendicular to the resonator's sensitive axis, hence isolating it from gravity sag and first-order stiffening effects. For in-orbit operations, the performance of the three sensors is expected to be comparable, since the resonant frequency of the Z sensor approaches those of the X and Y sensors under microgravity conditions, where flexure stiffening relaxes. Mechanisms explaining this phenomenon are discussed in Sec.~\ref{sec:z_results}.

\begin{table}[htbp]
\caption{Resonant Frequency and Quality Factor of the Z Resonator}
\label{tab:z_resonator_summary}
\centering
\begin{tabular}{lcc}
\hline
Configuration & $f_{0,z}$ (Hz) & $Q_z~(\times10^3)$ \\
\hline
In orbit (microgravity) & $11.07 \pm 0.15$ & $28.1 \pm 1.1$ \\
On ground ($1g$ load)   & $44.45 \pm 0.01$ & $6.65 \pm 0.13$ \\
\hline
\end{tabular}
\end{table}

The ring-down-measured $Q_z$ values in Table~\ref{tab:z_resonator_summary} both fall short of the $Q\sim10^{5}$ expected from independent anchoring, bulk, surface, and thermoelastic loss estimates, by roughly a factor of 4 in orbit and 15 on ground. Because the on-ground configuration introduces a large quadratic stiffness term absent in orbit, the more severe on-ground suppression is consistent with gravity-induced geometric nonlinearity. However, the in-orbit shortfall indicates that residual nonlinearity (namely the third-order stiffness term, which remains nonzero even without static loading) is itself sufficient to measurably broaden the resonance and suppress the observed $Q$ below the intrinsic loss budget.

A closely related mechanism has been reported in MEMS micromirror resonators, where amplitude- and pressure-dependent nonlinear damping measurably suppresses the ring-down-fitted $Q$ below intrinsic-loss estimates~\cite{Nabholz2018}. A more extreme example of the same underlying statistical effect (fluctuation broadening of the measured linewidth, distinct from genuine nonlinear damping) has also been reported in carbon-nanotube nanomechanical resonators, where thermally driven amplitude fluctuations alone suppressed the measured $Q$ by roughly two orders of magnitude below intrinsic-loss estimates~\cite{Barnard2012}; despite operating at GHz frequencies and nanogram-scale masses far different from our system, this result illustrates that fluctuation broadening can dominate ring-down-based $Q$ estimates across a wide range of physical scales. Distinguishing genuine nonlinear damping from fluctuation broadening typically requires fitting the ring-down envelope over restricted amplitude windows~\cite{Nabholz2018}; in our case, the seismic background near 11~Hz obscures the ring-down tail before the oscillation amplitude decays into the weakly nonlinear regime, precluding a reliable fit. A dedicated investigation of nonlinear damping in fused-silica resonators at these frequencies is left to future work.

\subsection{Optical readout and system characterization}
\label{sec:readout}

To measure displacements of the freely oscillating test mass, we utilize a dual heterodyne interferometer architecture that provides common-mode noise rejection, as described in our previous work~\cite{Hines2022, Zhang:22}. Each axis of the inertial sensor employs two frequency-shifted input beams ($f_1$, $f_2$) with a heterodyne beat frequency of $\sim$1~MHz, derived from a Coherent nonplanar ring oscillator (NPRO) laser at 1064~nm. The heterodyne light input is prepared with acousto-optic modulators (AOM), which impart a phonon-induced frequency shift to each beam.

\begin{figure}[htbp]
  \centering
  \includegraphics[width=\linewidth]{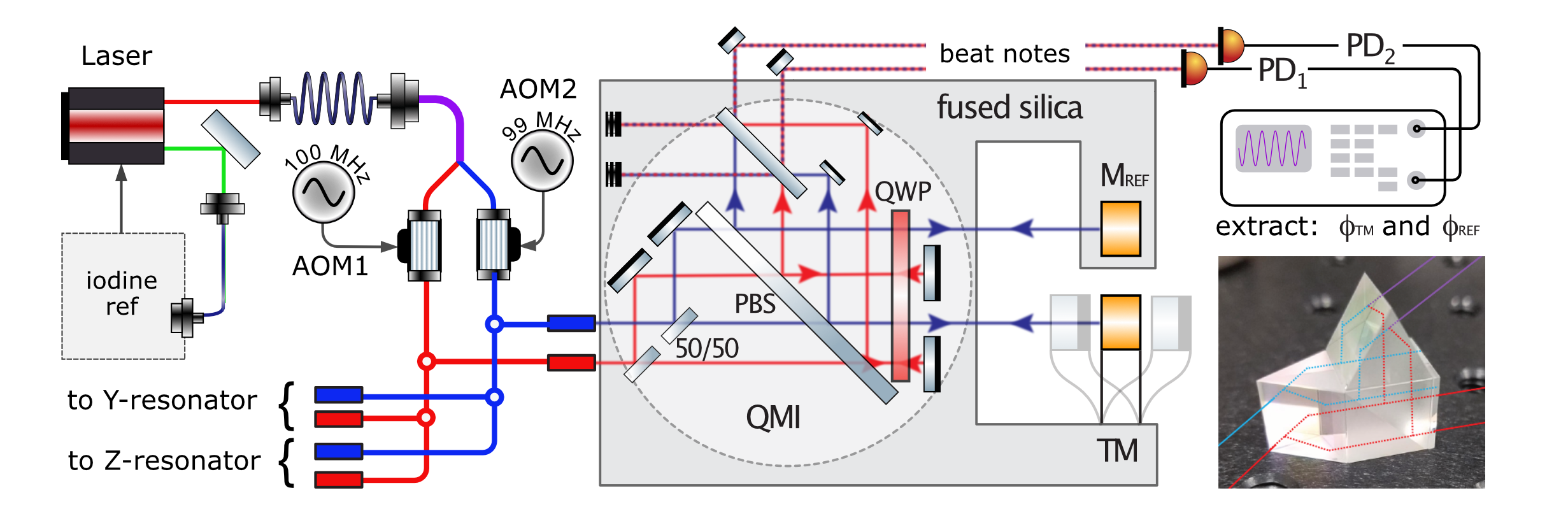}
  \caption{Dual heterodyne interferometer readout scheme. Input beams at frequencies $f_1$ and $f_2$ are split in a quasi-monolithic interferometer. One arm reflects from the moving test mass (TM) while the other reflects from a static reference mirror (RM) placed in close proximity. Both outputs are detected and sent to digital phasemeters for continuous phase tracking. The differential measurement provides common-mode rejection of environmental noise~\cite{Hines2022,Zhang:22}.}
  \label{fig:readout_scheme}
\end{figure}

Figure~\ref{fig:readout_scheme} shows the optical readout scheme in the triaxial sensor. The two frequency-shifted beams are generated via two fiber beam splitters, generating a relatively equal power level for each interferometer. The free-space optical segment utilizes a key quasi-monolithic interferometer (QMI) optic presented in detail in previous work~\cite{Zhang:22}, and allows for a dual interferometer setup with both a test-mass and reference interferometer. The QMI consists of a 50/50 beam splitter, polarizing beam splitter (PBS), and quarter-wave plate that split and recombine the input beams. Each interferometer utilizes a common arm reflecting from a highly reflective (HR) coating on the QMI. The test mass (TM) interferometer's signal arm reflects from the moving test mass mirror, while the reference (Ref) interferometer's signal arm reflects from a static reference mirror (RM) placed in close proximity to the test mass. Under optimal alignment of the test mass and reference mirrors, differential phase measurement between the TM and Ref interferometers rejects noise common to both interferometers. This includes but is not limited to laser frequency fluctuations, common temperature variations, vibrations, and fiber length changes while still preserving the inertial signal~\cite{Hines2022}.

For the triaxial system, three independent QMIs provide simultaneous readout of all three axes. As pictured in Fig.~\ref{fig:CAD_XYZ}, the QMIs in the XY axes lie adjacent to the test masses, whereas the test-mass beam for the Z axis is reflected 90 degrees through an aperture in the XY plate. Phase signals from each interferometer are continuously tracked using digital commercial phasemeters. The differential phase between test mass and reference interferometers yields the displacement measurement, which is then converted to acceleration using the known mechanical response of each resonator~\cite{Hines2020, Hines2022}.

We perform two types of measurements of the system in a vacuum chamber. The first set examines the performance of the triaxial sensor under nominal operation, with the test masses freely moving. A commercial Trillium 120 seismometer is placed in proximity to the triaxial sensor, acting as a reference to measure seismic signals. Post-corrections are then performed to simulate a seismic-free environment. The second set of measurements anchors the test masses, freezing motion, thereby giving us noise floors of the complete readout chain under a static test mass.

\begin{figure}[htbp]
    \centering
    \includegraphics[width=0.5\linewidth]{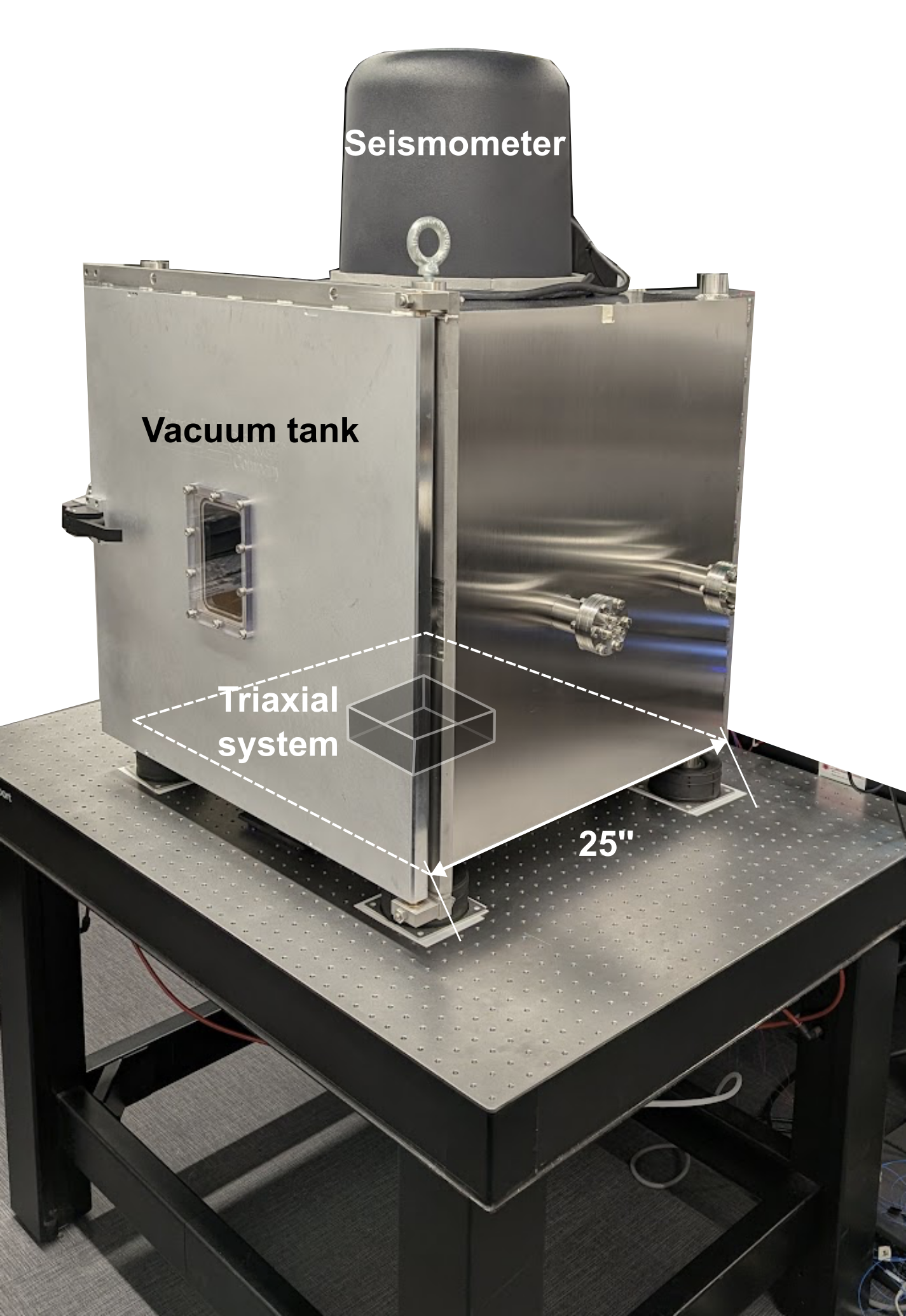}
    \caption{Vacuum tank housing the triaxial sensor, mounted on damped isolation legs on an optical table. The reference Trillium 120 seismometer (top) is mounted externally on the tank lid; the triaxial sensor sits at the base of the tank interior, separated from the seismometer by the full tank height.}
    \label{fig:vacuum_setup}
\end{figure}

The triaxial sensor is mounted at the base of a 25\,in $\times$ 25\,in $\times$ 25\,in vacuum tank, evacuated to $10^{-5}$\,mbar for all reported measurements. The tank is supported on damped isolation legs atop an optical table, and the reference Trillium 120 seismometer is mounted externally on top of the tank lid (Fig.~\ref{fig:vacuum_setup}). This arrangement introduces a fixed vertical and lateral offset between the two instruments' sensing points; as discussed in Sec.~\ref{sec:xy_performance}, this separation allows differential rotational motion between the two locations to appear as an uncorrelated residual after seismic subtraction, rather than reflecting a limitation of the sensor itself.

\subsection{Spectral analysis and seismic subtraction}
\label{sec:miso}

Amplitude spectral densities are computed using the log-frequency spectrum estimation method of Tr\"obs and Heinzel~\cite{trobs2006}, which averages multiple window lengths to give approximately uniform resolution on a logarithmic frequency axis while controlling spectral leakage. We use $J_{\mathrm{des}} = 500$ frequency bins for the final spectra, a Kaiser window with a peak sidelobe level of 200\,dB, and zero-order (mean) detrending of each segment. All spectral estimation and coherent subtraction is performed with the Python package \texttt{SpecKit}~\cite{speckit}.

Because the triaxial sensor and the reference Trillium 120 seismometer are digitized on independent, free-running clocks, their time series must first be synchronized before any cross-instrument comparison. We estimate the relative time shift by maximizing the coherence between the two channels within a frequency band of interest, and apply the resulting sub-sample delay using a Lagrange fractional-delay filter.

Seismic subtraction is performed using multiple-input/single-output (MISO) spectral analysis, applied independently to each axis of the triaxial sensor. For a given sensor axis, all three axes of the reference seismometer are treated as inputs to a linear, frequency-domain transfer function model whose predicted output is compared against the measured sensor channel; the coherent (transfer-function predicted) component of the seismic signal is then subtracted from the sensor spectrum at each frequency. Ambient lab pressure was also measured with a Adafruit BME680 environmental sensor and provided modest corrections to both in-plane sensors below 10 mHz. The residual, shown as the blue trace in Figs.~\ref{fig:accel_noise_xy} and~\ref{fig:accel_noise_z}, represents the portion of the sensor's measured motion that is not linearly coherent with the three-axis reference seismometer within the analysis band, and is bounded from below by the anchored (test-mass-fixed) readout noise floor (red trace).

\section{Triaxial sensor performance characterization}
\label{sec:performance}

We present acceleration measurements for all three axes of the triaxial sensor operating in vacuum at $10^{-5}$\,mbar. The XY axes exhibit spectral overlap with the reference seismometer across frequencies from 4\,mHz to 8\,Hz; in the following sections we discuss how our measurements validate on-ground performance. The Z axis exhibits higher noise on ground due to gravity-induced geometric stiffening effects mentioned in Sec.~\ref{sec:resonators}. The anchored test-mass measurements verify the low readout noise of the dual heterodyne interferometric approach. The X axis achieved noise floors of 60\,pico-g/$\sqrt{\mathrm{Hz}}$ at 1\,Hz with anchored test masses, demonstrating readout performance exceeding our previous single-axis readout noise levels~\cite{Hines2022}.

\subsection{X and Y sensors (in-plane)}
\label{sec:xy_performance}

\begin{figure}[htbp]
  \centering
  \includegraphics[width=\linewidth]{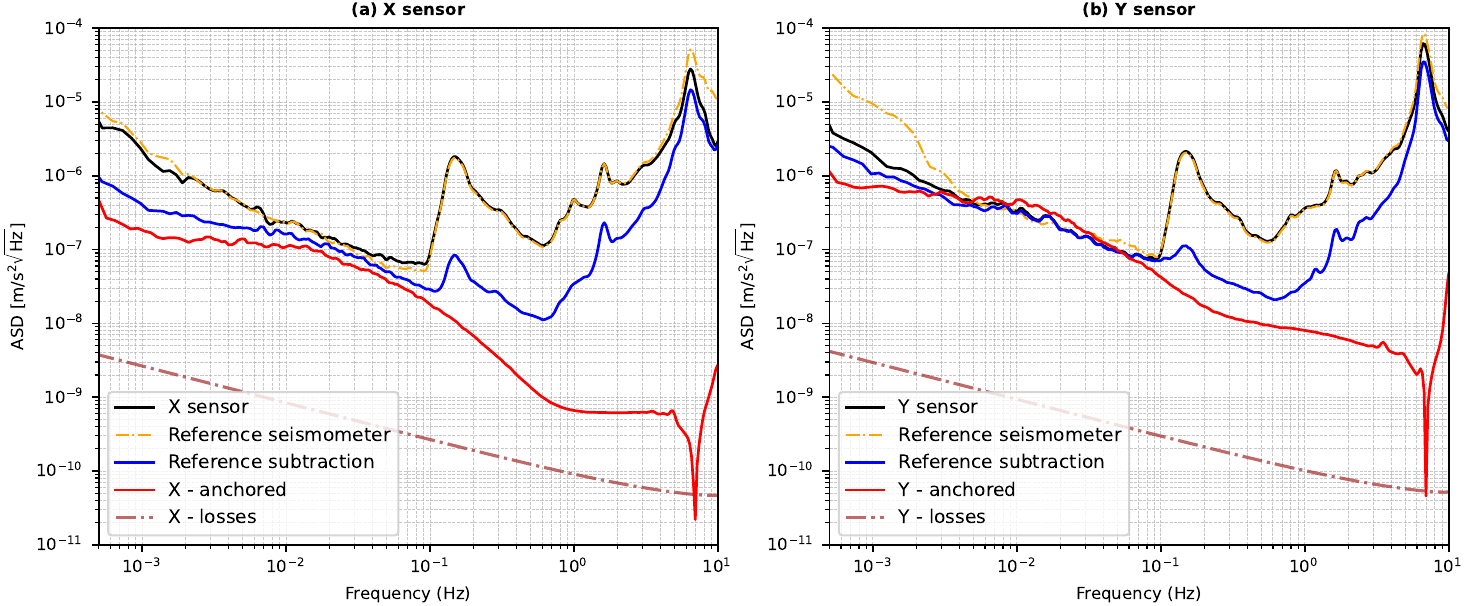}
  \caption{Performance of the X and Y sensors in the triaxial system shown by the black trace. The reference seismic measurement taken with an industrial seismometer is shown in yellow, highlighting agreement with the resonator. The noise after subtraction of this measured seismic signal is shown in blue, with still evident traces of a residual signal. The anchored noise is sketched in red to display the current readout noise levels. Expected thermal noise floors (dashed lines) are calculated from measured $Q$-factors using Eq.~(\ref{eq:thermal_noise}).}
  \label{fig:accel_noise_xy}
\end{figure}

The X-axis sensor shows spectral overlap with the signal recorded by the corresponding axis in the Trillium seismometer. After subtracting the seismic signal, we obtain the residual shown by the blue trace in Fig.~\ref{fig:accel_noise_xy}. We find minimal residual seismic signal below 0.1~Hz as there is good agreement with the anchored measurement noise floor, consistent with the anchored noise floor to within a factor of $\sim$3. The anchored test-mass measurements showcase low noise floors, down to 60\,pico-g/$\sqrt{\mathrm{Hz}}$ at 1\,Hz.

The discrepancy between the anchored noise floor and the subtracted result above 0.1\,Hz could be due to imperfect cancellation. Since the MISO model described in Sec.~\ref{sec:miso} assumes a linear, time-invariant relationship between the reference seismometer and the sensor, it cannot capture non-stationary or rotationally induced coupling; for instance, tilt-to-acceleration cross-coupling arising because the triaxial system and the seismometer occupy different positions in space and therefore sense differential rotational motion not present in the three-axis MISO input set. Such effects would appear in the residual as apparently uncorrelated noise even though their physical origin is still the ambient seismic field. For a true subtraction, rotations need to be independently monitored. Additionally, subtractions are made easier if one starts with a quieter environment. As such, to test the true potential of the triaxial sensor, we will place it in facilities with better seismic isolation. Future iterations will also include replacing photodiodes with quadrant photodiodes to better estimate tilt-to-length effects, as the test mass may experience small angular rotations during oscillations.

Similar performance is found with the Y resonator as well, with agreement from 4~mHz to 8~Hz with the measured seismic signal. As with the X axis, the corrected data for the Y axis still shows a residual microseism. There is closer agreement between the anchored and the corrected traces. However, this is due to the higher readout noise in the Y axis. This is seen clearly in Fig.~\ref{fig:accel_noise_xy}, and highlights the improvements possible for the Y axis due to the major commonalities shared with the X axis.

We measure higher noise levels for the anchored measurements with the Y sensor, ranging from a factor of 2 at 0.1\,Hz to a factor of 10 at 1\,Hz. We identify two candidate contributors to this excess. First, post-measurement inspection revealed that the Y-axis anchoring fails to fully seat against the test mass, allowing residual low-level test-mass motion to couple into the nominally anchored measurement; this would raise the apparent anchored noise floor without reflecting the sensor's true intrinsic readout noise. Second, the Y interferometer exhibited lower fringe visibility than the X interferometer, consistent with imperfect beam alignment, which independently raises the phase readout noise floor. Disentangling these contributions requires re-characterization with a corrected anchoring fixture and improved Y-interferometer alignment, which we defer to future work. Given that the X and Y resonators are of identical design, we expect such a re-characterization to bring the anchored Y noise floor close to that measured on the X axis.

\subsection{Z sensor and additional noise characterization}
\label{sec:z_results}

\begin{figure}[htbp]
  \centering
  \includegraphics[width=0.7\linewidth]{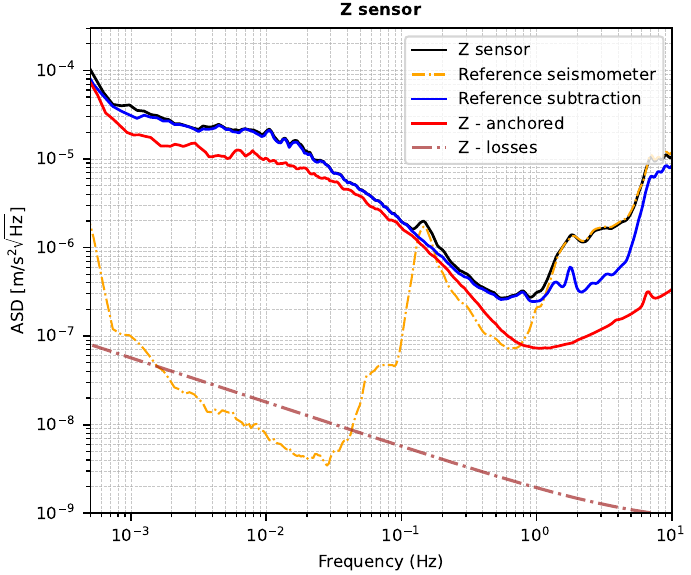}
  \caption{Performance of the Z sensor in the integrated triaxial system (black). Owing to the higher flexure stiffness under $1g$ loading, the Z resonator is less sensitive to the microseism than the X and Y axes, but still resolves the peak near 0.1~Hz and vibrations above 1~Hz, as evidenced by the overlap with the reference seismometer (orange dash-dot). The seismic-subtracted residual is shown in blue and the anchored test-mass readout floor in red. The dash-dotted brown line is the expected thermal-noise floor calculated from the measured $Q$-factor.}
  \label{fig:accel_noise_z}
\end{figure}

In the integrated triaxial sensor, we identify two independent, quantitatively separable noise contributions limiting the Z-axis on-ground performance: gravity-induced stiffening of the flexures, and a non-optimal interferometer geometry inherited from the integrated wafer layout. The first raises the resonant frequency from 11.07~Hz to 44.45~Hz under $1g$ loading and is absent in orbit (Table~\ref{tab:z_resonator_summary}), reducing the Z-axis acceleration response by $(44.45/6.88)^2 \approx 42$ relative to the X and Y axes. The second arises because the XY wafer lacks the clearance to accommodate two down-propagating beams that would allow us to place the Z reference mirror next to the test mass; instead, the reference mirror sits on the XY wafer, increasing the optical path difference (OPD), leading to higher laser frequency noise coupling. Even so, the Z sensor resolves the tip of the microseism near 0.1~Hz (black trace, Fig.~\ref{fig:accel_noise_z}) and agrees with the reference seismometer (orange dash-dot) above 1~Hz, confirming that both instruments track the same ground motion; the peak vanishes after seismic subtraction (blue), confirming its seismic origin.

The Z resonator geometry itself was redesigned from an earlier out-of-plane design that suffered static test-mass tilt and nonlinear motion from laterally shifted flexures. We adopted a bridge-like geometry with four symmetric flexures, matched to the X/Y etch direction, which eliminates measurable tilt and yields markedly better surface quality in the flexures.

\begin{figure}[htbp]
  \centering
  \includegraphics[width=\linewidth]{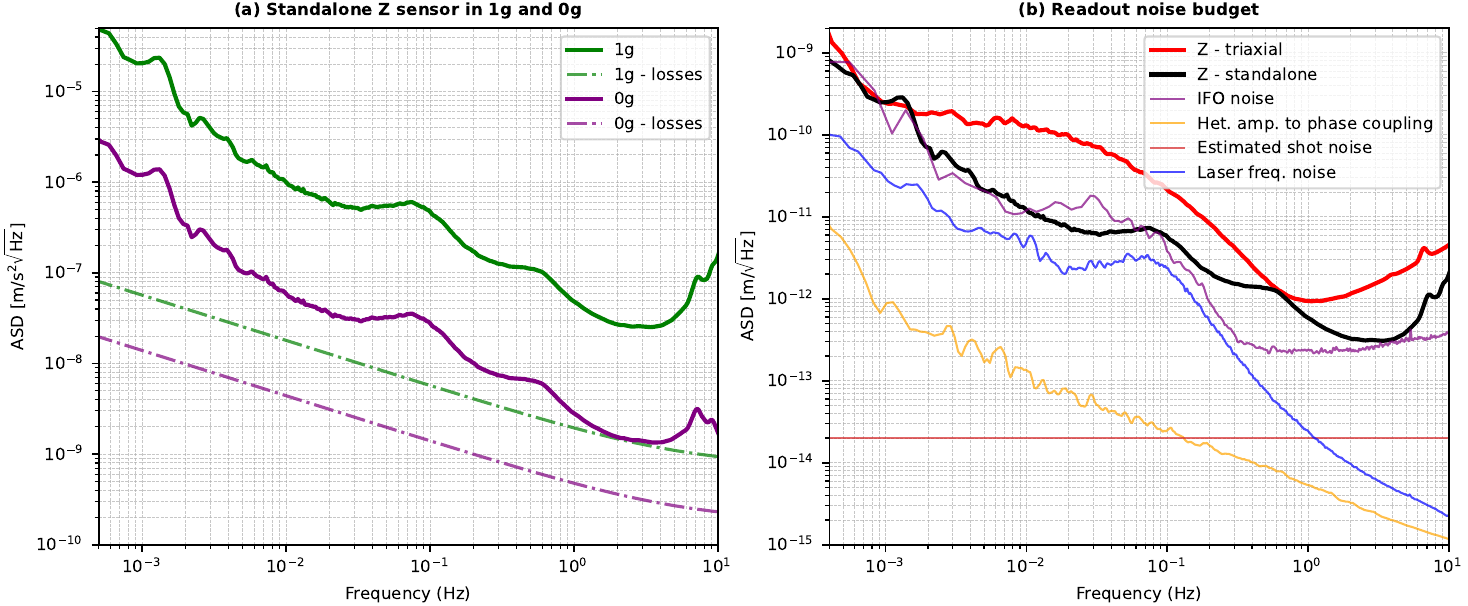}
  \caption{(a) Acceleration performance of the standalone Z-axis sensor under nominal $1g$ loading (green) and the same measurement rescaled to the $0g$ resonant frequency and quality factor (purple); dash-dot lines show the corresponding thermal-noise floors at $1g$ and $0g$. Under the projected $0g$ condition the noise is a factor of $\sim$5 above thermal at 2~Hz and $\sim$10 at 10~mHz. (b) Readout noise budget of the standalone configuration. The optimized standalone geometry (black) improves on the integrated triaxial readout (red) by roughly an order of magnitude at 10~mHz. Contributing terms are the interferometer/QMI noise (purple), scaled RIN (tan), laser frequency noise (blue), and the estimated shot-noise floor (red, horizontal); the free-running measurement is limited by laser frequency noise near 0.1~Hz.}
  \label{fig:z_readout}
\end{figure}

To isolate and address the interferometer-geometry contribution, we characterized a standalone Z-axis testbed with an optimized optical layout: the XY wafer was replaced by an aluminum plate with a large aperture that passes both the test-mass (TM) and reference (RM) beams, allowing us to place the reference mirror directly adjacent to the test mass as in the X and Y axes. This shortens the differential path length and improves common-mode rejection. The result is a clean, order-of-magnitude improvement: the standalone configuration (black, Fig.~\ref{fig:z_readout}(b)) outperforms the integrated triaxial readout (red) by roughly an order of magnitude at 10~mHz and remains consistently lower above 2~mHz. This isolates the interferometer geometry as the dominant readout-noise contribution in the integrated system and shows that a larger aperture in future XY wafer designs recovers Z-axis readout performance comparable to the X and Y axes, with the improvement demonstrated here carrying directly over to the integrated instrument.

The remaining contributions to the standalone floor are decomposed in Fig.~\ref{fig:z_readout}(b). The measurement used a free-running Coherent NPRO laser; with a residual OPD of 3~mm between the test-mass and reference interferometers, it is limited by laser frequency noise (blue) near 0.1~Hz. The interferometer (QMI) noise (purple) was characterized separately, in a configuration where a single static mirror served as the common target for both the test-mass and reference interferometers. This term is limited by temperature fluctuations at the lowest frequencies, by beam jitter driving alignment fluctuations at mid frequencies, and by parasitic interference from secondary beams at the highest frequencies. Certain fluctuations in the heterodyne amplitude and in the measured phase share common origins, so the amplitude channel serves as a diagnostic for phase noise. Modulating the AOM drive at 10~Hz and comparing the resulting peaks in the phasemeter amplitude and phase outputs gives a coupling of 200~pm per fractional amplitude fluctuation. The measured relative amplitude fluctuation spectrum scaled by this coefficient is plotted in tan. The coefficient is determined at a single modulation frequency and applied across the band, and measurements at additional modulation frequencies would be required to establish its true frequency dependence. The laser power was high enough such that the measurement is not shot-noise limited (red).

To recover the intrinsic microgravity behavior of the resonator, we rotated the standalone Z plate by 90$^\circ$ so that the test-mass motion is perpendicular to gravity, removing the gravity-induced stiffening. In this effective $0g$ configuration the resonant frequency drops to $f_{0,z} = 11.07$~Hz with $Q_z = 28100$ (Table~\ref{tab:z_resonator_summary}; see Sec.~\ref{sec:resonators} for the associated $Q$ uncertainty). Rescaling the standalone $1g$ measurement to these values yields the projected $0g$ sensitivity shown by the purple trace in Fig.~\ref{fig:z_readout}(a): a factor of $\sim$5 above the thermal floor at 2~Hz and $\sim$10 at 10~mHz. Given that the $0g$ resonant frequency approaches those of the X and Y sensors, this projection indicates comparable performance across all three axes in orbit.

\section{Conclusion}
\label{sec:conclusion}

We have demonstrated a compact triaxial optomechanical accelerometer achieving simultaneous readout of all three orthogonal axes. Laboratory measurements of the X axis achieved noise floors of 60~pico-g/$\sqrt{\mathrm{Hz}}$ at 1~Hz, while the Y axis exhibited higher noise (factor of 2--10) due to imperfect anchoring during characterization measurements. Both axes showed agreement with a commercial seismometer across frequencies from 4~mHz to 8~Hz, validating performance for ground-based seismometry applications. The anchored test-mass measurements showcase the low readout noise of the dual heterodyne interferometric system, validating the optomechanical approach for low-frequency inertial sensing.

The performance of the Z axis is currently limited by static gravitational loading during ground testing. The resonant frequency increases from 11.07~Hz under microgravity conditions to 44.45~Hz under $1g$ load, reducing the acceleration response by a factor of approximately 41 relative to the X and Y
sensors in our laboratory environment. However, measurements based on a horizontal configuration indicate that the Z axis will achieve comparable sensitivity to the X and Y axes in a microgravity setting.

This work demonstrates a complete triaxial optomechanical accelerometer occupying a compact footprint of 156$\times$156$\times$65\,mm$^3$ with two axes achieving seismometer-grade sensitivity and a third axis whose microgravity performance is well characterized. This triaxial system demonstrates high sensitivity, compact size, and simultaneous three-axis readout, enabling applications in both ground-based seismometry and space-based precision measurements.


\section*{Funding}
National Aeronautics and Space Administration Instrument Incubator Program Award 3055360.

\section*{Acknowledgment}
The authors thank Pengzhuo Wang for data on interferometric noise sources, Miguel Dovale for developing the spectral analysis code, Jackson Dahn for helpful discussions on machining, and Brina Martinez for coating the test-mass mirrors. The authors also thank the rest of the Laboratory of Space Systems and Optomechanics (LASSO) members at the University of Arizona's James C. Wyant College of Optical Sciences for their support.

\section*{Disclosures}
The authors declare no conflicts of interest.

\section*{Data availability}
Data underlying the results presented in this paper are not publicly available at this time but may be obtained from the authors upon reasonable request.


\bibliographystyle{unsrt}
\bibliography{references}

\end{document}